%% file: paper.tex
\documentclass[sigconf,nonacm]{acmart}

\usepackage{listings}
\lstdefinelanguage[ECMAScript2015]{JavaScript}[]{JavaScript}{
  morekeywords=[1]{await, async, case, catch, class, const, default, do,
    enum, export, extends, finally, from, implements, import, instanceof,
    let, static, super, switch, throw, try},
  morestring=[b]` 
}
\lstdefinelanguage{JavaScript}{
  morekeywords=[1]{break, continue, delete, else, for, function, if, in,
    new, return, this, typeof, var, void, while, with},
  morekeywords=[2]{false, null, true, boolean, number, undefined,
    Array, Boolean, Date, Math, Number, String, Object},
  morekeywords=[3]{eval, parseInt, parseFloat, escape, unescape},
  sensitive,
  morecomment=[s]{/*}{*/},
  morecomment=[l]//,
  morecomment=[s]{/**}{*/}, 
  morestring=[b]',
  morestring=[b]"
}[keywords, comments, strings]
\lstalias[]{ES6}[ECMAScript2015]{JavaScript}
\definecolor{mediumgray}{rgb}{0.3, 0.4, 0.4}
\definecolor{mediumblue}{rgb}{0.0, 0.0, 0.8}
\definecolor{forestgreen}{rgb}{0.13, 0.55, 0.13}
\definecolor{darkviolet}{rgb}{0.58, 0.0, 0.83}
\definecolor{royalblue}{rgb}{0.25, 0.41, 0.88}
\definecolor{crimson}{rgb}{0.86, 0.8, 0.24}
\lstdefinestyle{JSES6Base}{
  backgroundcolor=\color{white},
  basicstyle=\ttfamily,
  breakatwhitespace=false,
  breaklines=false,
  captionpos=b,
  columns=fullflexible,
  commentstyle=\color{mediumgray}\upshape,
  emph={},
  emphstyle=\color{crimson},
  extendedchars=true,  
  fontadjust=true,
  frame=single,
  identifierstyle=\color{black},
  keepspaces=true,
  keywordstyle=\color{mediumblue},
  keywordstyle={[2]\color{darkviolet}},
  keywordstyle={[3]\color{royalblue}},
  numbers=left,
  numbersep=5pt,
  numberstyle=\tiny\color{black},
  rulecolor=\color{black},
  showlines=true,
  showspaces=false,
  showstringspaces=false,
  showtabs=false,
  stringstyle=\color{forestgreen},
  tabsize=2,
  title=\lstname,
  upquote=true  
}
\lstdefinestyle{JavaScript}{
  language=JavaScript,
  style=JSES6Base
}
\lstdefinestyle{ES6}{
  language=ES6,
  style=JSES6Base
}

\usepackage{algorithm}
\usepackage{algpseudocode}
\algrenewcommand\algorithmicrequire{\textbf{Input:}}
\makeatletter
\newcommand{\StatexIndent}[1][3]{%
  \setlength\@tempdima{\algorithmicindent}%
  \Statex\hskip\dimexpr#1\@tempdima\relax}
\makeatother

\usepackage[nolist]{acronym}
\usepackage{subcaption}

\AtBeginDocument{%
  \providecommand\BibTeX{{%
    \normalfont B\kern-0.5em{\scshape i\kern-0.25em b}\kern-0.8em\TeX}}}

\begin{document}

\title{Quantifying Semantic Query Similarity for Automated Linear SQL Grading: A Graph-based Approach}

\author{Leo K{\"o}berlein}
\affiliation{%
  \institution{Friedrich-Alexander-University}
  \city{Erlangen}
  \country{Germany}
}
\email{leo.koeberlein@fau.de}

\author{Dominik Probst}
\affiliation{%
  \institution{Friedrich-Alexander-University}
  \city{Erlangen}
  \country{Germany}
}
\email{dominik.probst@fau.de}

\author{Richard Lenz}
\affiliation{%
  \institution{Friedrich-Alexander-University}
  \city{Erlangen}
  \country{Germany}
}
\email{richard.lenz@fau.de}


\begin{abstract}
  Quantifying the semantic similarity between database queries is a critical challenge with broad applications, ranging from query log analysis to automated educational assessment of SQL skills. Traditional methods often rely solely on syntactic comparisons or are limited to checking for semantic equivalence.

  This paper introduces a novel graph-based approach to measure the semantic dissimilarity between SQL queries. Queries are represented as nodes in an implicit graph, while the transitions between nodes are called edits, which are weighted by semantic dissimilarity. We employ shortest path algorithms to identify the lowest-cost edit sequence between two given queries, thereby defining a quantifiable measure of semantic distance.

  A prototype implementation of this technique has been evaluated through an empirical study, which strongly suggests that our method provides more accurate and comprehensible grading compared to existing techniques. Moreover, the results indicate that our approach comes close to the quality of manual grading, making it a robust tool for diverse database query comparison tasks.
\end{abstract}





\maketitle

\begin{acronym}[DBMS]
  \acro{ast}[AST]{Abstract Syntax Tree}
  \acro{dbms}[DBMS]{Database Management System}
  \acro{iff}[iff]{if and only if}
  \acro{sql}[SQL]{Structured Query Language}
  \acro{uml}[UML]{Unified Modeling Language}
\end{acronym}
\acused{sql}

\input{src/introduction}
\input{src/related-works}
\input{src/concept}
\input{src/evaluation}
\input{src/future-work}
\input{src/conclusion}

\begin{acks}
  We thank the participants of our survey and all proofreaders.
\end{acks}

\newpage

\bibliographystyle{ACM-Reference-Format}
\bibliography{literatur}

\newpage
\appendix

\section{Uniform Cost Search}\label{sec:uniform-cost-search}


\begin{figure}[htb]
  \centering
  \includegraphics[width=0.45\linewidth]{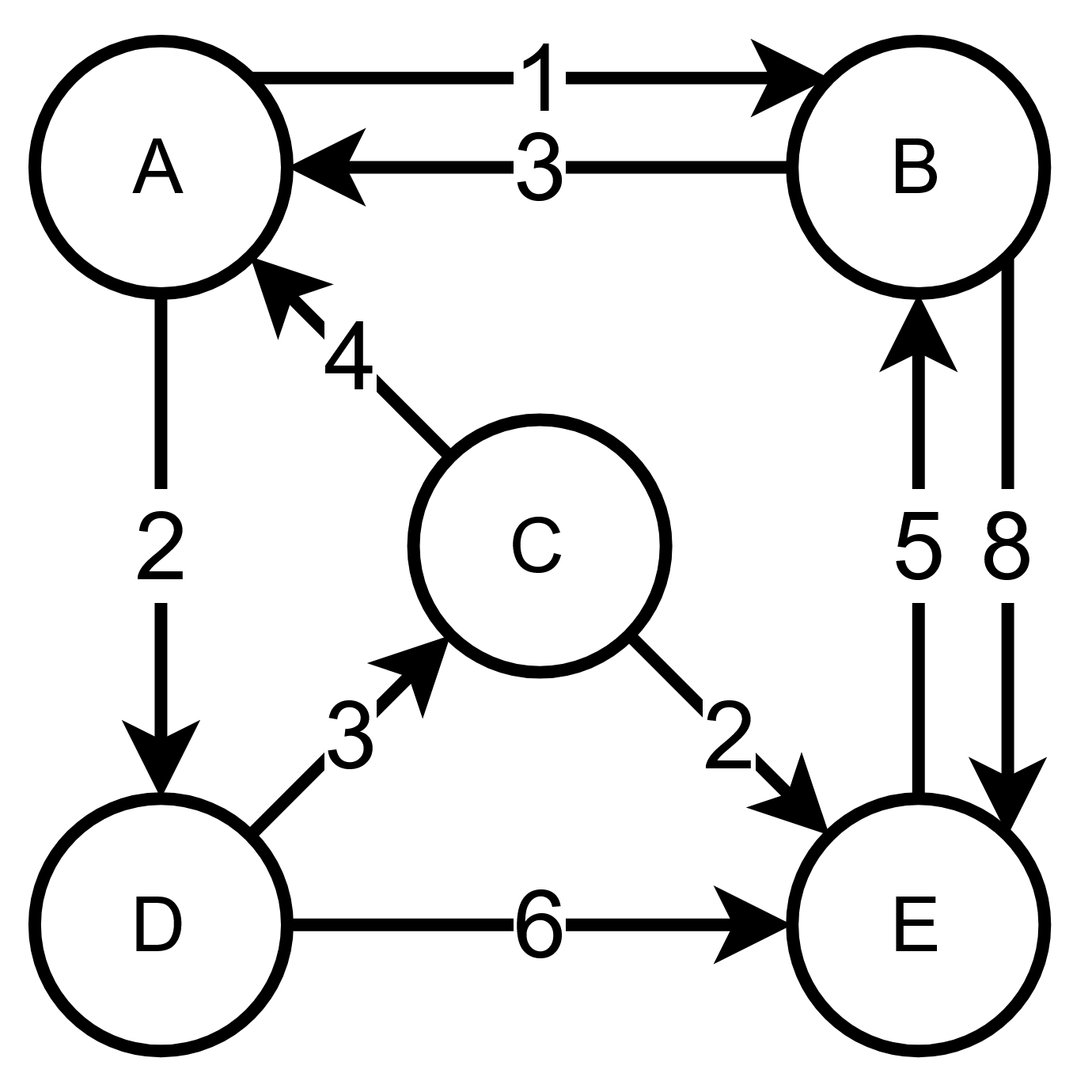}
  \caption{Simple graph consisting of five nodes and nine weighted edges.}
  \Description{A graph consisting of nodes A-E connected via weighted edges.}
  \label{fig:simple-graph}
\end{figure}

The shortest path problem is to find the path from a start to a destination node, for which the sum of costs of its edges is the lowest.
For example, in the graph in \autoref{fig:simple-graph}, the shortest path from node $A$ to node $E$ is $A \to D \to C \to E$.

Dijkstra's algorithm is a famous algorithm for solving the shortest path problem.
However, in its commonly taught form, it cannot be applied to implicit graphs because it would need to know all nodes of the graph right from the beginning. \cite{Felner.2011}
Instead, uniform cost search can be used, which is shown in \hyperref[alg:ucs]{Algorithm \ref*{alg:ucs}}.

It uses a priority queue called \texttt{UNVISITED}, which is initialized with the start node \texttt{s} at its distance of \texttt{0}.
While this queue is not empty, the node \texttt{u} with minimal distance is extracted.
Because it has the minimal distance, there can be no other path to it with a smaller one (assuming non-negative edge costs).
So it is added to the set of visited nodes, called \texttt{VISITED}, whose distances will not change anymore.
Then, its neighboring nodes, along with the cost of the respective edge, are determined via the neighbor function of the implicit graph.
For each neighbor, it is checked whether it has already been visited before, in which case it is skipped because the shortest path to it was already found.
Otherwise, it is either added to the priority queue with a tentative distance, in case it was undiscovered before.
Or, in case it was already discovered, its tentative distance is updated to a potentially smaller value by using \texttt{u} as an intermediate step along the path.

Every node, apart from the start node, is at first undiscovered.
It is discovered but still unvisited, as soon as one of its neighbors becomes visited, at which point it receives a tentative distance.
If more of its neighbors are visited before itself, its tentative distance may get updated to a smaller value.
When it is extracted from the priority queue and becomes visited, its tentative distance is the smallest one of all unvisited nodes, so there can be no shorter path to it using another unvisited node.
Its tentative distance becomes its final distance.

\begin{figure}[htb]
  \centering
  \includegraphics[width=\linewidth]{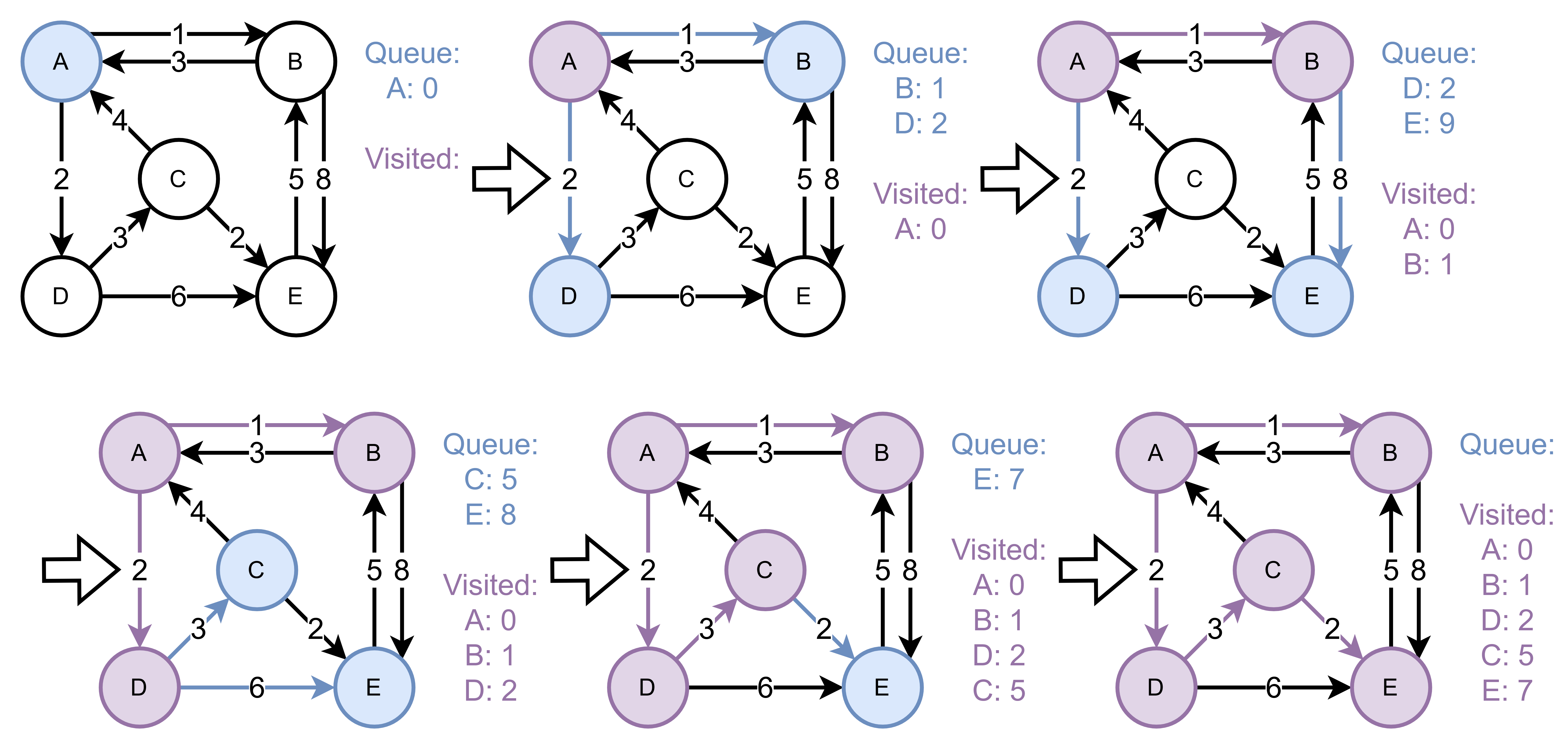}
  \caption{The steps of applying uniform cost search with start node $A$ to the simple graph.}
  \Description{[A], [A, B], [A, B, D], [A, B, D, C], [A, B, D, C, E]}
  \label{fig:simple-graph-ucs}
\end{figure}

\autoref{fig:simple-graph-ucs} shows the steps of applying uniform cost search to the graph from \autoref{fig:simple-graph}, starting at node $A$.
The queue of discovered, but unvisited nodes with their tentative distances and the respective paths in the graph are colored blue.
The set of visited nodes with their final distances and the respective paths in the graph are colored purple.
Notice how the tentative distance of node $E$, as well as the path to it, change multiple times as shorter alternatives are discovered.

In the form described above, the algorithm builds a tree of shortest paths from the start node as root to every other node in the graph.
But often only the shortest path to one specific destination node is of interest.
In this case, the algorithm can be stopped as soon as this destination becomes visited, because its distance and path will not change anymore.
This is again assuming all edge costs are non-negative.
For example, if node $C$ from above was the destination, the algorithm could have stopped after step five.

\begin{algorithm}[htb]
  \caption{Uniform Cost Search}\label{alg:ucs}
  \begin{algorithmic}[1]
    \Require Start Node \texttt{s}
    \State \texttt{dist[s] $\gets$ 0}
    \State \texttt{UNVISITED.insert(s,\ 0)}
    \While{\texttt{UNVISITED $\neq$ $\emptyset$}}
    \State \texttt{u $\gets$ UNVISITED.extract\_min()}
    \State \texttt{VISITED.insert(u)}
    \For{(Node, Cost) \texttt{(n,\ c) $\in$ Neighbors(u)}}
    \If{\texttt{n $\notin$ VISITED}}
    \If{\texttt{n $\notin$ UNVISITED}}
    \State \texttt{dist[n] $\gets$ dist[u] + c}
    \State \texttt{UNVISITED.insert(n,\ dist[n])}
    \Else
    \State \texttt{dist[n] $\gets$ min(dist[n],\ dist[u] + c)}
    \State \texttt{UNVISITED.update(n,\ dist[n])}
    \EndIf
    \EndIf
    \EndFor
    \EndWhile
  \end{algorithmic}
\end{algorithm}

\end{document}

%% file: src/introduction.tex
\section{introduction}\label{sec:introduction}




\label{sec:motivation}

The problem of comparing \ac{sql} queries has received some attention in the past, both theoretical and practical \cite{Chandra.2021,Aho.1979,Chu.2018,Zhou.2019}.
These studies generally focus on trying to answer the question of whether two given queries are equivalent or, more generally, whether the result of one is a superset of the other, the so-called containment problem.
However, especially in the area of teaching \ac{sql}, there is a need to answer a different kind of comparison problem.

To allow for partial grading, a finer differentiation than just a binary scale is required.
Not only is it of interest whether a student's query and a reference solution are completely equivalent, but if they are not, their similarity needs to be quantified.

It is also beneficial for a student's learning process if meaningful feedback is provided.
This means explaining why their solution is not (fully) correct, e.g., what steps would be necessary to make it correct.
And if it is correct but looks different from the reference solution, it is desirable to be able to explain why it is still correct.

These tasks are still usually done by hand, especially for exams.
This is not only time-consuming, but also error-prone, as the severity of errors can be subjective and vary between correctors or subtle mistakes can be missed.
An automated solution to these problems is therefore required.

It is likely that an approach for measuring a quantifiable semantic distance between queries would also be beneficial in other application areas. For example, the search for similar queries in query logs as described in \cite{Wahl.2017} or in \cite{Bonifati.2019} could be significantly improved.

\section{Goals}\label{sec:goals}

The goals of this paper can be summarized as follows:

\begin{description}
    \item[Quantified semantic similarity:]\label{item:goal-quantify}
          Two \ac{sql} queries are to be compared.
          The result of the comparison should not be equivalence, but instead similarity, quantified on a linear scale.
          The similarity should express how much the queries differ semantically, not just syntactically, to make it more useful for grading.

    \item[Meaningful feedback:]\label{item:goal-feedback}
          It is desirable to have detailed semantic feedback.
          There should be explanations as to either why the two queries were found to be equivalent, or why they were not equivalent and what would be required to change that.
          Such feedback is helpful in the development of the implementation, but it can also help students to understand their mistakes in exercises and justify the reasoning behind grades in exams.

    \item[Guaranteed result:]\label{item:goal-result}
          The method of comparison should always produce a result.
          Given that query equivalence is generally undecidable \cite{Chu.2017b, Abiteboul.1995}, this is not trivial.

    \item[Unrestricted input:]\label{item:goal-input}
          Arbitrary \ac{sql} queries should be processable.
          Some, mostly theoretical, approaches try to circumvent this undecidability by restricting themselves to small subsets of \ac{sql}.
          Our comparison technique should not be restricted in this way.

    \item[Configurability and extensibility:]\label{item:goal-configure}
          The system should be configurable and extensible.
          Different exercises or problems require different points of focus.
          It should be easy to take this into account and to adapt the comparison.
\end{description}








%% file: src/related-works.tex
\section{Related Works}\label{sec:related-works}

Other publications on comparing \ac{sql} queries can be roughly categorized into three groups:
Those that do not compare the queries themselves, but rather the tuples returned when they are executed on a sample database, which is called dynamic analysis.
Others, that compare the queries themselves using static analysis, in order to determine whether or not they are semantically equivalent.
And finally those that also use static analysis, but continue to check how similar they are in case they are not completely equivalent.

\subsection{Dynamic analysis}\label{sec:comparing-execution-results}

Dynamic analysis is a popular way of comparing \ac{sql} queries. \cite{Bhangdiya.2015, Chandra.2015, Dominguez.2019, Kleiner.2013, Nalintippayawong.2017, Prior.2004, Soler.2006}
For example, given a reference solution R and a student-devised solution S, both are considered semantically equivalent if the combined query in \autoref{lst:compare-statement} does not return any tuples.

\begin{lstlisting}[language=SQL, caption={Basic \acs{sql} statement for comparing queries via their execution results.}, captionpos=b, label=lst:compare-statement]
    (S UNION R) EXCEPT (S INTERSECT R)
\end{lstlisting}

The advantage is the conceptual simplicity:\\
The reference solution is an executable query, i.e., it can be executed on the database.
For the student's query to be a correct solution to the problem, it must be executable and (always) return the same tuples.
And since it is executed on a real \ac{dbms}, parsing and error checking is automatically handled and the query behaves as it would in the real world.

Currently the most advanced take on this is the XData system by \citet{Bhangdiya.2015} and \citet{Chandra.2015} developed at IIT Bombay.
Their goal is to improve the precision of dynamic analysis by generating special test databases for each individual query.
These will produce differing results in the presence of typical errors associated with certain query constructs.

The disadvantages of this approach are significant:\\
Dynamic analysis can only prove that queries \emph{are not} equivalent, but not that they \emph{are}.
For example, there may be cases where there are subtle differences that only lead to different results for very specific combinations of data.
If these are missing from the test databases, they will go undetected, leading to false positives.
The risk of missing such cases is quite high. Even \citet{Chandra.2015} acknowledge this problem and only strive to detect common errors, while some query constructs are not even fully supported by the data generation \cite{Chandra.2019}.

Secondly, dynamic analysis itself only gives a binary result.
Trying to estimate similarity based on the fraction of matching output tuples works poorly because tiny differences, e.g., in the selection, can lead to greatly diverging results.

Finally, there is no meaningful feedback.
When queries are found not to be equivalent, there is no explicit explanation as to why they produced different outputs.

\subsection{Static analysis for equivalence}\label{sec:comparing-for-equivalence}

A promising approach is static analysis, which compares the queries themselves.
Given that query equivalence is generally undecidable \cite{Chu.2017b, Abiteboul.1995}, there are two avenues to pursue in order to uphold theoretical correctness:

One way is to consider only a subset of \ac{sql}, for example conjunctive queries \cite[chapter 4]{Abiteboul.1995}, on which equivalence is decidable.
Possible decision techniques are those of \citet{Aho.1979}, \citet{Chu.2018} and \citet{Zhou.2019}.
This is mainly of theoretical relevance, since in practice, a wider range of queries is required.

The other way is to have fewer restrictions on the queries, but this runs the risk of not being able to prove or disprove their equivalence in some cases.
Cosette by \citet{Chu.2017b} tries to strike a balance.
It is a powerful tool and has been used to prove and disprove many equivalence transformations used in execution plan optimizers.
However, there are still some features that are not yet supported, such as the aggregation functions \texttt{AVG}, \texttt{MIN} and \texttt{MAX}, or the \texttt{ORDER-BY} clause.

Also in this category resides an approach by \citet{Dollinger.2011}.
It is similar to ours, in that it uses equivalence transformation patterns to try and turn one query into the other.
The main differences are that it only checks for equivalence and focuses on high-level transformations involving subqueries, while seemingly neglecting lower-level equivalences, such as the distributive law.

All these techniques have major drawbacks:\\
Either there are restrictions on the query constructs supported, or there is no guaranteed answer.
Moreover, they still only determine equivalence, not quantified similarity.

\subsection{Static analysis for similarity}\label{sec:comparing-for-similarity}

To address the need for comparing queries beyond pure equivalence, methods for determining their similarity have emerged.

A simple approach is to use string similarity metrics to compare the descriptions of the reference solution and the student-devised query.
\citet{Stajduhar.2015} combine multiple different such metrics (including absolute length difference, Levenshtein distance and Euclidean word frequency distance) and feed them into a logistic regression model, which is trained on a large number of manually assigned scores.
This has been tested with some success, but it is obvious that it will only give approximate results.

The problem of partial correctness is being addressed by \citet{Fabijanic.2020} by decomposing the query into clauses (e.g. \texttt{SELECT}, \texttt{FROM}, \texttt{WHERE}, ...).
Then, a series of successive scoring rules is applied, each of which checks whether each clause is the same as the corresponding clause in the reference, and deducts a fixed score if it is not.
As a measure of similarity, this is better than a binary result, but still a rather rough distribution.

A more detailed and mathematical method is presented by \citet{Panni.2020}, combining the two ideas to some extent.
The queries are split into clauses, but for each clause, an individual score is calculated using mathematical formulas.
These formulas are tailored to each clause and take into account the exact number of individual matching and mismatching elements.
The individual scores are then combined into a weighted sum, giving an accurate result.

All mentioned approaches have the disadvantage that, apart from the basic structure of a query, they completely ignore the semantics of \ac{sql}.
The location of an error is not as important as the logical step it represents, so weighting by component is of little use.
Moreover, there are numerous possibilities of having two semantically equivalent queries, that are syntactically different.
The only way to produce correct results would be to provide a large number of queries, that are semantically equivalent to the original reference solution, which is unrealistic.

\citet{Wang.2020} first check all student-developed queries for semantic equivalence to the reference solution, and then use these as if they were reference solutions in calculating the partial scores for all non-equivalent queries.
The initial equivalence check is performed using dynamic analysis (see \autoref{sec:comparing-execution-results}).
Partial marks are generally assigned by parsing the query into an \ac{ast} and determining the tree edit distance.
This idea is similar to \citet{Dollinger.2011} and our approach.
If the query description cannot be parsed into an AST, string similarity is used instead, as in \cite{Stajduhar.2015}.
The calculation of partial marks using the tree-edit distance or string similarity ignores the semantics of \ac{sql} and is only approximate.
In addition, dynamic analysis is not sufficient as an initial test for semantic equivalence, mainly because of the risk of false positives, as explained in \autoref{sec:comparing-execution-results}.

The XData system (see \autoref{sec:comparing-execution-results}) has been extended by \citeauthor{Chandra.2021} \cite{Chandra.2016,Chandra.2019b,Chandra.2019,Chandra.2021,Chandra.2022} to be able to generate partial marks as well.
As in the previous approach, a student-developed query is first checked against the reference solution by comparing their execution results.
Then an edit distance is calculated. However, unlike a basic tree edit distance, the semantics of \ac{sql} are taken into account:
Small edits, e.g. adding/removing/replacing an attribute or changing a \texttt{JOIN}-type, are applied successively to the student's query.
These edits only add elements that are actually present in the sample solution and remove those that are not.
After each edit, equivalence to the sample solution is checked by applying a normalisation, called canonicalisation, which consists of a series of confluent equivalence transformations.
As a result, if the query is semantically equivalent despite syntactic differences, it is correctly identified as such.
This approach still carries the risk of false positives during the initial equivalence check (see \autoref{sec:comparing-execution-results}) and can only process queries that are executable.
Since the idea behind the partial marking is quite similar to the one presented in this paper, deeper comparisons will be introduced throughout \autoref{sec:concept}.
It will be referred to as "\citeauthor{Chandra.2021}".

Most approaches introduced in this section suffer from a lack of meaningful feedback.
The last two approaches are able to provide such feedback if the queries aren't equivalent, but, due to the preceding dynamic analyis-step, unable if they are.

%% file: src/concept.tex
\section{Concept}\label{sec:concept}

This paper presents a new method of comparing \ac{sql} queries,
that achieves the goals set out in \autoref{sec:goals} without the shortcomings of previous approaches shown in \autoref{sec:related-works}.

\subsection{Definitions}\label{sec:definitions}

The following definitions clarify the terminology used throughout this paper:

\begin{definition}[Parsable]
    A query is parsable \ac{iff} it follows the \ac{sql} syntax and can therefore be parsed into an \acf{ast} representation.
\end{definition}

\begin{definition}[Executable]
    A query is executable in the context of a given database schema, \ac{iff} it is parsable and can be executed on a database of said schema (without throwing errors).
\end{definition}

\begin{definition}[Syntactically equivalent]
    Two queries are syntactically equivalent \ac{iff} their \ac{ast} representations are exactly the same.
\end{definition}

\begin{definition}[Semantically equivalent]
    Two queries are semantically equivalent in the context of a given database schema, \ac{iff} they return exactly the same result when executed on an arbitrary but fixed database of said schema.
\end{definition}

While syntactically equivalent queries are always semantically equivalent, there may be queries that are semantically equivalent but not syntactically.
For example, suppose the two tables \texttt{students} and \texttt{teachers} both have a column \texttt{id}. Then the queries in \autoref{lst:query-1-of-semantically-equivalent-pair} and \autoref{lst:query-2-of-semantically-equivalent-pair} are semantically equivalent and should be recognized as such, even though they are syntactically different.

\begin{lstlisting}[language=SQL, caption={Query 1 of semantically equivalent query pair.}, captionpos=b, label=lst:query-1-of-semantically-equivalent-pair]
    SELECT *
    FROM students JOIN teachers
    ON students.id = teachers.id
\end{lstlisting}

\begin{lstlisting}[language=SQL, caption={Query 2 of semantically equivalent query pair.}, captionpos=b, label=lst:query-2-of-semantically-equivalent-pair]
    SELECT *
    FROM students, teachers
    WHERE students.id = teachers.id
\end{lstlisting}

This example also shows that in order to determine semantic equivalence (and hence similarity), parts of a query cannot simply be compared individually without the context of the rest of the query.
Consider also the queries in \autoref{lst:query-1-of-possibly-equivalent-pair} and \autoref{lst:query-2-of-possibly-equivalent-pair}.

\begin{lstlisting}[language=SQL, caption={Query 1 of possibly equivalent pair.}, captionpos=b, label=lst:query-1-of-possibly-equivalent-pair]
    SELECT students.id
    FROM students, teachers
\end{lstlisting}

\begin{lstlisting}[language=SQL, caption={Query 2 of possibly equivalent pair.}, captionpos=b, label=lst:query-2-of-possibly-equivalent-pair]
    SELECT id
    FROM students, teachers
\end{lstlisting}

For the same schema as before, these two queries are not equivalent.
The second one is not even executable, because the column name \texttt{id} is ambiguous.
But if only the table \texttt{students} had a column called \texttt{id}, the queries would be semantically equivalent (and executable).
This means that semantic equivalence (and executability) depends not only on the queries, but also the context of the database schema the queries are executed on.

\subsection{Idea}\label{sec:idea}


Queries are nodes in a graph.
The graph is implicit and edits on these queries are neighbor functions which, when applied to a query/ node, generate neighboring nodes. 
Each edit 
has a (non-negative) cost associated with it, representing the error or dissimilarity it creates.
Given this graph, a shortest path algorithm can be used to find the lowest-cost edit sequence from a start node to a destination node.
The combined weight of this edit sequence quantifies, how dissimilar the start and target query are.

\begin{figure}[htb]
    \centering
    \includegraphics[width=0.8\linewidth]{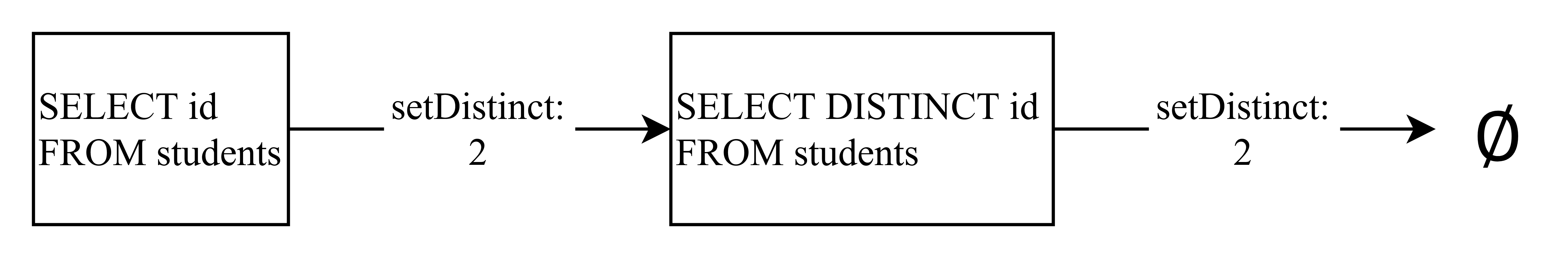}
    \caption{Two subsequent applications of the edit with cost 2 called \texttt{setDistinct} on the start query.}
    \Description{When trying to apply \texttt{setDistinct} a second time, the result is the empty set.}
    \label{fig:set-distinct}
\end{figure}

Suppose we have a table \texttt{students} with columns \texttt{id}, \texttt{name}, and \texttt{age}, and a query based on this schema, shown in \autoref{lst:start-query}, serving as a starting point.
Assume there is an edit called \texttt{setDistinct} with a cost of 2, which sets the \texttt{DISTINCT} declaration on a query.
Applying this edit to the start query creates the neighbor in \autoref{lst:start-query-distinct}.
Attempting to apply the same edit to this neighbor will not create a new node, as it is already \texttt{DISTINCT}.
This is visualized in \autoref{fig:set-distinct}.

\begin{lstlisting}[language=SQL, caption={Start query.}, captionpos=b, label=lst:start-query]
    SELECT id
    FROM students
\end{lstlisting}

\begin{lstlisting}[language=SQL, caption={\texttt{DISTINCT} neighbor of start query.}, captionpos=b, label=lst:start-query-distinct]
    SELECT DISTINCT id
    FROM students
\end{lstlisting}

\begin{figure}[htb]
    \centering
    \includegraphics[width=\linewidth]{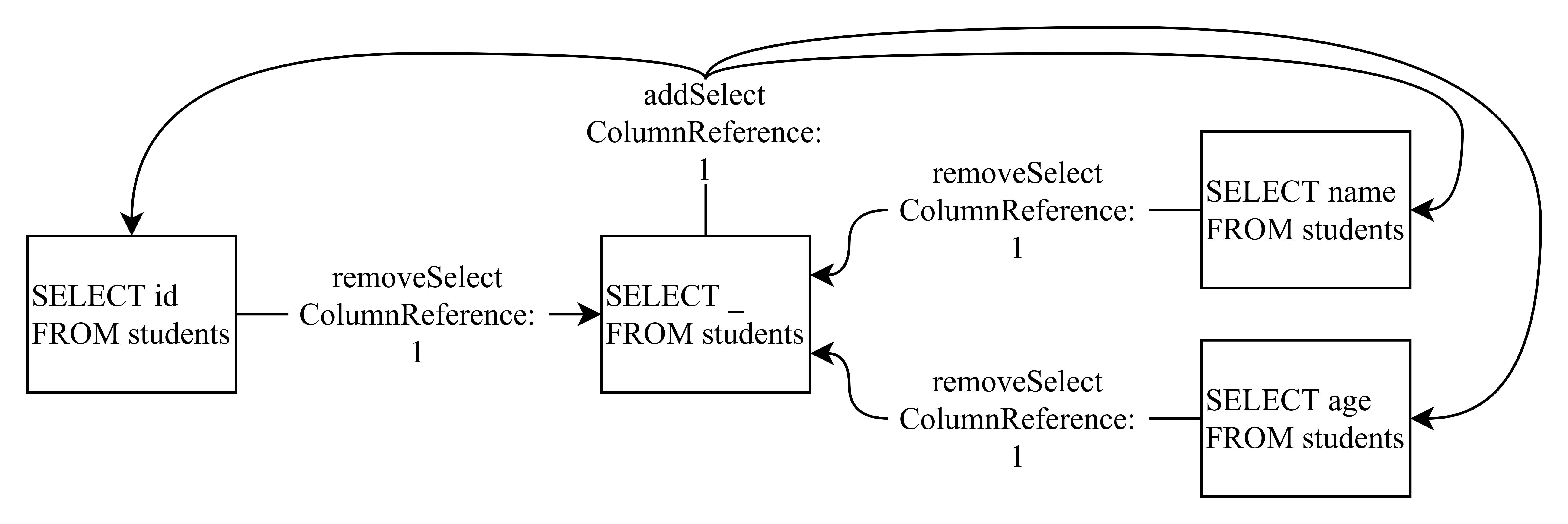}
    \caption{Multiple neighbors from one application of \texttt{addSelectColumnReference}.}
    \Description{Applying \texttt{addSelectColumnReference} to the query yields three neighbors, one for each column that is available via the \texttt{FROM}-clause.}
    \label{fig:add-remove-column-reference}
\end{figure}

Consider an edit \texttt{addSelectColumnReference} with cost 1, which adds a column  reference to the expression of a \texttt{SELECT} element, and another edit \texttt{removeSelectColumnReference} with cost 1, which removes one.
Applying the former to the start query will not work because it has only one \texttt{SELECT} element, and its expression is already occupied by a column reference to \texttt{id}.
Applying the latter will produce a new neighbor that still has a \texttt{SELECT} element, but now with an empty expression, symbolised as \texttt{\_}.
Doing the same on the new neighbor does not work because there are no more column references to remove.
But now \texttt{addSelectColumnReference} can be applied, resulting in three neighbors, one for each available column.
One is the start query again.
And on the other two, \texttt{removeSelectColumnReference} can be applied in the same way.
This is visualized in \autoref{fig:add-remove-column-reference}.

\begin{figure}[htb]
    \centering
    \includegraphics[width=\linewidth]{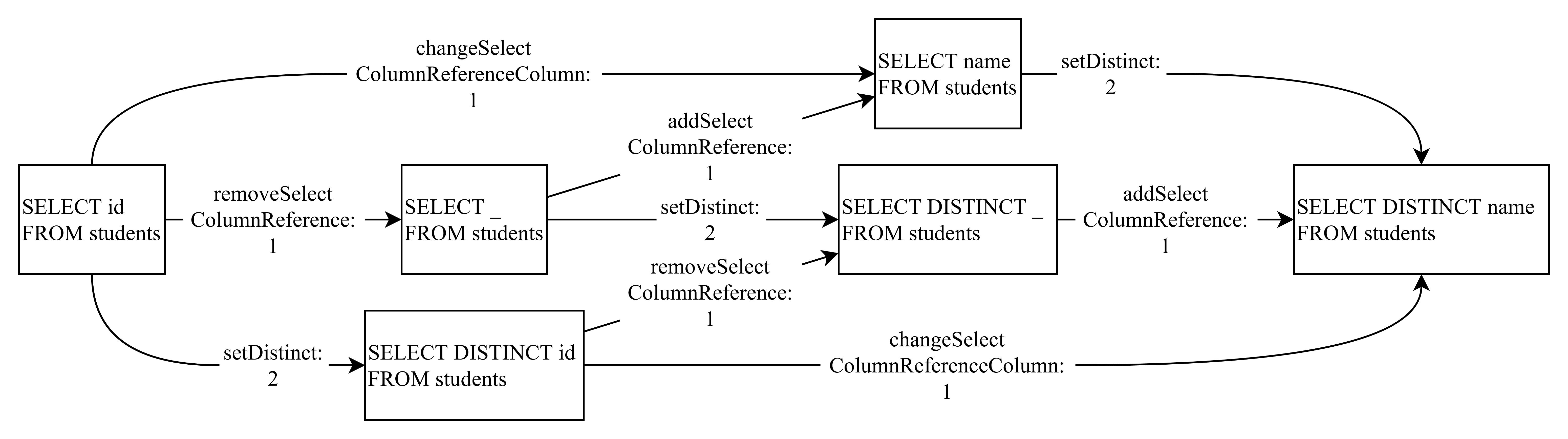}
    \caption{Possible paths from the start query to the destination query.}
    \Description{Depending on the first edit applied, different paths with different sums of costs open up.}
    \label{fig:possible-paths}
\end{figure}

The outer three nodes can be reached by deleting and re-adding a column reference.
This is unnecessarily expensive, so there is another edit \texttt{changeSelectColumnReferenceColumn} with cost 1, which changes the referenced column.
Finally, suppose we want to compare the destination query in \autoref{lst:destination-query} with the start query.
Combining all of the above, there are several possible paths between them.
The relevant ones are shown in \autoref{fig:possible-paths}, but for the sake of simplicity nothing strongly divergent.

\begin{lstlisting}[language=SQL, caption={Destination query.}, captionpos=b, label=lst:destination-query]
    SELECT DISTINCT name
    FROM students
\end{lstlisting}

Based on this graph, a shortest path algorithm can be used to find the cheapest edit sequence between the start and destination.
In the example above, there are two equally cheap ones, consisting of \texttt{changeSelectColumnReferenceColumn} and \texttt{setDistinct} in any order.
The sum of their costs is 3, so this is the quantified semantic dissimilarity between the start and target query.

\input{src/concept/nodes}

\input{src/concept/edges}

\input{src/concept/algorithm}

\input{src/concept/termination}

\subsection{Grading}\label{sec:result}

The result of the shortest path algorithm is the smallest total distance between the start and destination query.
To use it for grading, this value is subtracted from the maximum possible points for that task.
Since the distance can be large, the points are lower-bounded to 0.
If the cost of the edits, and therefore the total distance, is of a different order of magnitude than the maximum points for the tasks, a constant scaling factor can be applied before subtraction.
The grading formula is:
$$points = max(maxPoints - distance * scale, 0)$$

This approach can also be used to determine a reasonable value for this maximum number of points for an assignment.
Since incomplete \acp{ast} can be processed, the "empty \ac{ast}" can be given as a start.
The total distance to the reference solution quantifies the work to build the target from scratch, and is therefore a good "difficulty" measure.


%% file: src/concept/nodes.tex

\subsection{Nodes}\label{sec:nodes}


Nodes represent one of the two principal components of a graph. 
In order to define them, syntactic and semantic distinctions and the matter of including non-executable queries are relevant.

\subsubsection{Syntactic and semantic differences}\label{sec:syntactic-and-semantic-differences}

The relevance of certain syntactic and semantic differences varies depending on the focus of teaching, or whether the comparison context is related to teaching at all.

Suppose a table \texttt{students} has a column \texttt{id} as its primary key.
Then the \texttt{DISTINCT} declaration could be removed from the query in \autoref{lst:query-with-unnecessary-distinct} and the query would still be semantically equivalent to before.
From an objective point of view, this is an equivalence transformation and should have no cost associated with it.
From a pedagogical point of view, it could be argued that a student should be penalised for using an unnecessary declaration, so there should be a cost associated with it.

\begin{lstlisting}[language=SQL, caption={\acs{sql} query with unnecessary \texttt{DISTINCT}-declaration.}, captionpos=b, label=lst:query-with-unnecessary-distinct]
    SELECT DISTINCT id
    FROM students
\end{lstlisting}

This becomes even more important when an assignment is specifically intended to teach a particular syntactic construct.
For example, the \texttt{WITH} clause can be eliminated without changing the semantics by replacing all references to it within the query with its contents.
Applying this transformation to the query in \autoref{lst:query-with-with-clause} results in the semantically equivalent query in \autoref{lst:query-with-eliminated-with-clause}.
Normally there should be no cost associated with this change.
However, if the task description explicitly asks students to use a \texttt{WITH} clause, it will be necessary to penalise a student who does not use it.

\begin{lstlisting}[language=SQL, caption={\acs{sql} query with a \texttt{WITH}-clause.}, captionpos=b, label=lst:query-with-with-clause]
    WITH subquery(averageAge) AS (
        SELECT AVG(age) AS averageAge 
        FROM students)
    SELECT id
    FROM students, subquery
    WHERE students.age > subquery.averageAge
\end{lstlisting}

\begin{lstlisting}[language=SQL, caption={\acs{sql} query with eliminated \texttt{WITH}-clause.}, captionpos=b, label=lst:query-with-eliminated-with-clause]
    SELECT id
    FROM students, (
        SELECT AVG(age) AS averageAge 
        FROM students) AS subquery
    WHERE students.age > subquery.averageAge
\end{lstlisting}


\subsubsection{Non-executability}\label{sec:syntactic-and-semantic-incorrectness}

An important question is how to handle queries that are not executable.

It is sensible to allow non-executable queries as intermediate steps along the path.
For example, assuming that the tables \texttt{students} and \texttt{teachers} both have a column \texttt{id} and the former also has a column \texttt{name}, the query in \autoref{lst:semantically-incorrect-intermediate-step} is non-executable, but is a possible intermediate step between the queries in \autoref{lst:query-before-intermediate-step} and \autoref{lst:query-after-intermediate-step}.

\begin{lstlisting}[language=SQL, caption={Query before intermediate step.}, captionpos=b, label=lst:query-before-intermediate-step]
    SELECT students.name
    FROM students, teachers
\end{lstlisting}

\begin{lstlisting}[language=SQL, caption={Non-executable intermediate step.}, captionpos=b, label=lst:semantically-incorrect-intermediate-step]
    SELECT students.name, id
    FROM students, teachers
\end{lstlisting}

\begin{lstlisting}[language=SQL, caption={Query after intermediate step.}, captionpos=b, label=lst:query-after-intermediate-step]
    SELECT students.name, students.id
    FROM students, teachers
\end{lstlisting}

This step could be skipped by using a more complex edit that detects that the column reference is ambiguous without specifying the table, and therefore adds it immediately.

If students produce faulty queries, they could not be dealt with if non-executable queries were not allowed as nodes.
The same holds true for incomplete queries, like the one in \autoref{lst:incomplete-query}, where students grasp the general structure but are only missing minor parts of the query.

\begin{lstlisting}[language=SQL, caption={Incomplete query missing an expression inside the aggregation function.}, captionpos=b, label=lst:incomplete-query]
    SELECT AVG(   )
    FROM students
\end{lstlisting}

The graph should be able to contain non-executable and even incomplete queries, making for less complicated edits and higher flexibility internally, as well as allowing for more freedom in the input.

\subsubsection{Node definition and comparison}\label{sec:node-definition-and-comparison}

The definition of a node in the graph is closely tied to specifying a comparison operation, as it dictates whether two given queries are distinct or part of the same node. We define a node as follows:

A node is one particular \ac{ast}, which may have missing subcomponents or unset attributes.
Nodes are compared directly, i.e., the position and type of each (sub)component and the value of each attribute must match.
Two given queries are considered semantically equivalent \ac{iff} there is a path with a cumulative cost of 0 between them.

%

Our definition stands in contrast to that used by \citeauthor{Chandra.2021}, as they consider a node to be a class of semantically equivalent queries. We are therefore able to distinguish between syntactic and semantic differences, as discussed in \autoref{sec:syntactic-and-semantic-differences}, whereas they are not. In addition, \citeauthor{Chandra.2021} only consider executable queries \cite{Chandra.2019}, which means that many queries produced by students cannot be processed by their system like described in \autoref{sec:syntactic-and-semantic-incorrectness}.



%% file: src/concept/edges.tex
\subsection{Edits}\label{sec:edges}


In graphs, besides nodes, there are usually edges. But in our approach, they are implicitly represented by edits. These are differentiated into a fundamental set of atomic edits and the semantics-aware shortcut edits.

\subsubsection{Edits vs. edges}\label{sec:edits-vs-edges}

As mentioned in \autoref{sec:idea}, the graph is implicit.
Thus, edges between nodes are not explicitly given, but there is a neighbor function that returns the neighbors of a given node.
The edits are a set of such neighbor functions.
They can be combined into a single one by applying them all and returning the union of their results as a single set of neighbors.

\begin{figure}[htb]
	\centering
	\includegraphics[width=\linewidth]{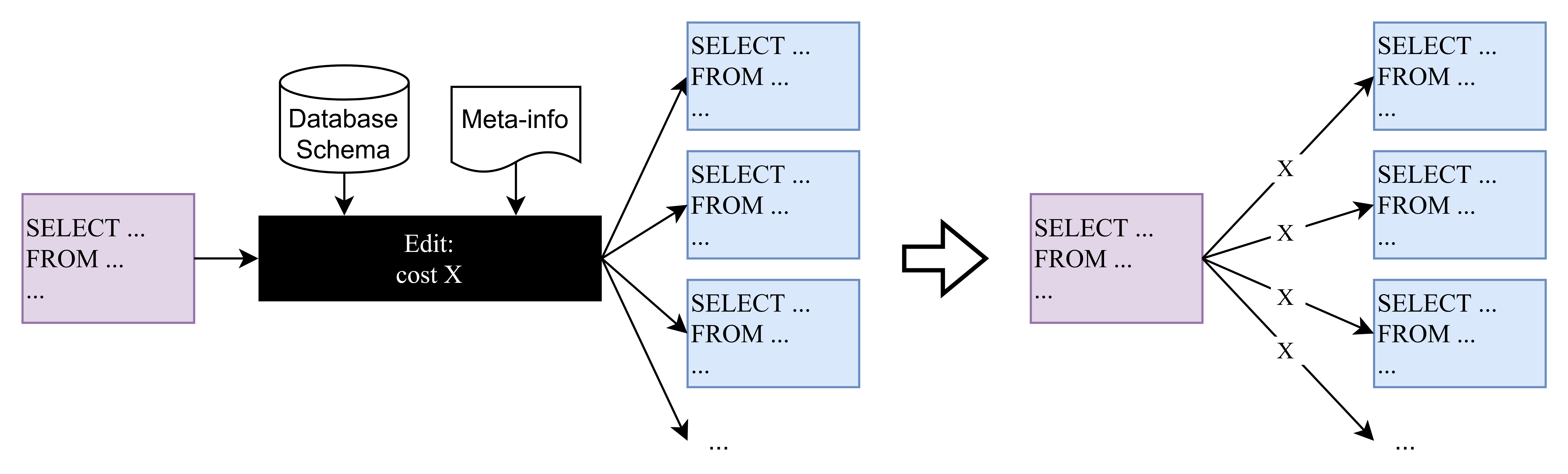}
	\caption{One edit representing multiple outgoing edges.}
	\Description{Applying one edit to a query can yield multiple results, so this edit represents multiple edges, one to each result.}
	\label{fig:edits-vs-edges}
\end{figure}

An edit is a 1-to-n function that takes an \ac{ast} as input and returns a set of different \acp{ast}.
This set can be empty if the conditions for applying the edit are not met by the input.
For example, a query cannot be made \texttt{DISTINCT} if it already is.
An edit can represent multiple outgoing edges.
This is visualized by \autoref{fig:edits-vs-edges}.
The graph is generally directed.
It is possible to make it undirected, but this requires a lot of work to ensure that each edit has a corresponding counter-edit.

An edit might need the database schema as context to preserve semantic equivalence (see \autoref{sec:definitions}).
Certain information about the destination query is extracted into meta-info.
These additional inputs can be seen in \autoref{fig:edits-vs-edges} as well.

Each edit has a fixed, non-negative cost.
This cost does not change at runtime. If it is necessary to weight an edit differently in different parts of the query, it must be replaced by multiple part-specific edits.
The same principle can be applied to other desired distinctions.

\subsubsection{Types of Edits}\label{sec:edit-set-construction}

We distinguish several different kinds of edits:

\textbf{Atomic edits} are the fundamental edits that make a small, atomic change to the \ac{ast} of a query. For example, for every (sub-) component, there are respective add- and remove-edits, and for every attribute, there are respective set- and unset-edits.
They are responsible for connecting every node in the graph, so they are inherently important to ensure termination.
They do not have any conditions, apart from there actually being an empty space to fill, existing component to remove or variable to set/unset.
Generally, they change the query semantics and therefore have a cost greater 0.
This cost must represent the greatest semantic difference they can cause, even if this over-estimates the distance in other cases.

\begin{figure}[htb]
	\centering
	\includegraphics[width=\linewidth]{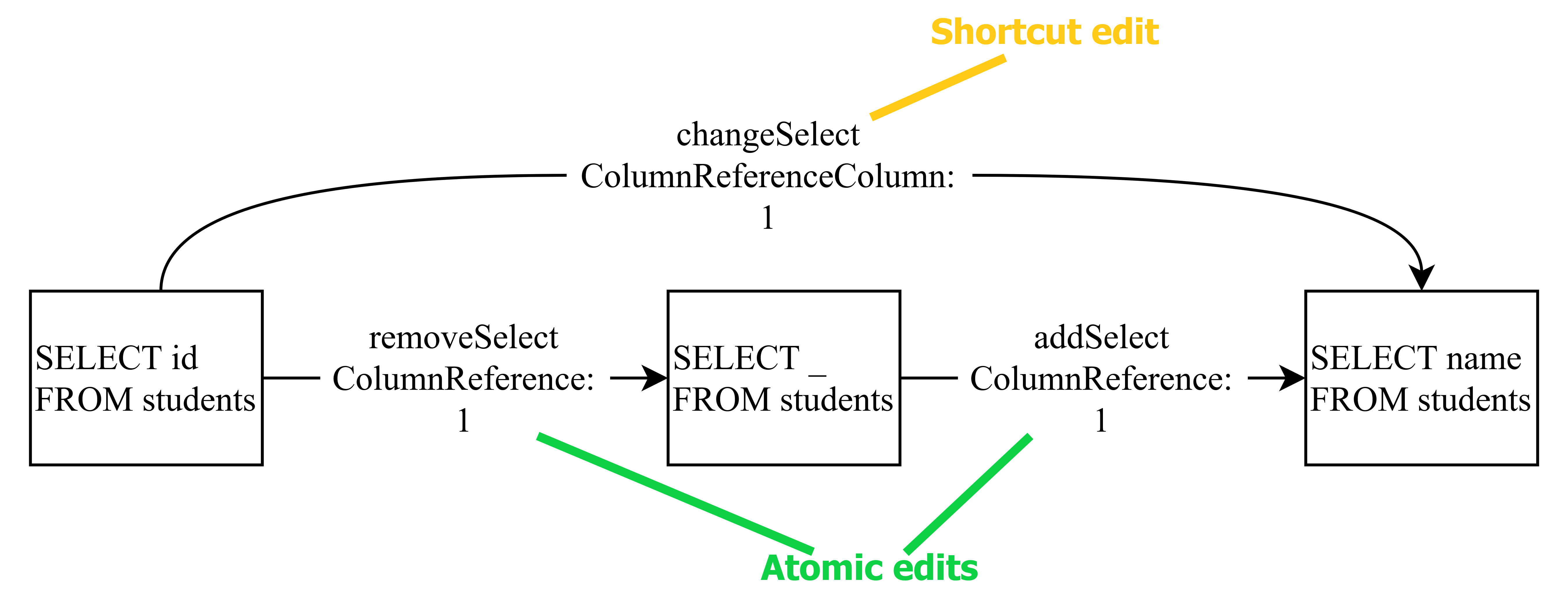} 
	\caption{Two atomic edits with combined cost 2 vs. one shortcut edit with cost 1.}
	\Description{What a shortcut edit does can also be done via (composited) atomic edits, but it is cheaper.}
	\label{fig:atomic-vs-shortcut}
\end{figure}

\textbf{Shortcut edits} rely on certain conditions in order to be applicable but in turn perform transformations, that would require a combination of multiple atomic operations, for a cost, that is lower than the cumulated cost of the corresponding atomic edits.
This is visualized in \autoref{fig:atomic-vs-shortcut}.

\textbf{Horizontal edits} are an intuitive subset of shortcut edits. Their defining characteristic is that they only swap (sub)components or change values, but don't add/remove or set/unset. Examples include the shortcut edit from \autoref{fig:atomic-vs-shortcut}, swapping the positions of two \texttt{SELECT}-elements, and the associative law.

\textbf{Equivalence edits} are edits with a cost of 0. They connect those queries, that are to be considered semantically equivalent, and are mostly shortcut edits, but there might also be certain atomic edits among them, depending on the exact implementation of the \ac{ast} and configuration.
These edits can never be fully exhaustive, because otherwise, this approach would be able to decide query equivalence, which is impossible. \cite{Chu.2017b, Abiteboul.1995}

The exact costs of edits depend on the goals of the comparison, e.g., different priorities regarding certain syntactic or semantic differences, as pointed out in \autoref{sec:syntactic-and-semantic-differences}.

\subsubsection{Finetuning of Edits}\label{sec:edit-set-improvement}

A base set of edits can be expanded for each use case with task-specific shortcut edits.

When introducing new shortcut edits, attention should be paid to the intended purpose:
If we, for example, consider resolving an \texttt{INNER JOIN} by creating a cross product with an associated \texttt{WHERE} clause, it's quite clear that a semantically equivalent query arises. A shortcut edit with a cost of 0, i.e., an equivalence edit, could be introduced.
Such an edit should not apply to every \texttt{JOIN}, though. In the case of a \texttt{LEFT JOIN}, there is no equivalence, and the costs for a potential shortcut edit would need to be higher than 0.

\begin{figure}[htb]
	\centering
	\includegraphics[width=\linewidth]{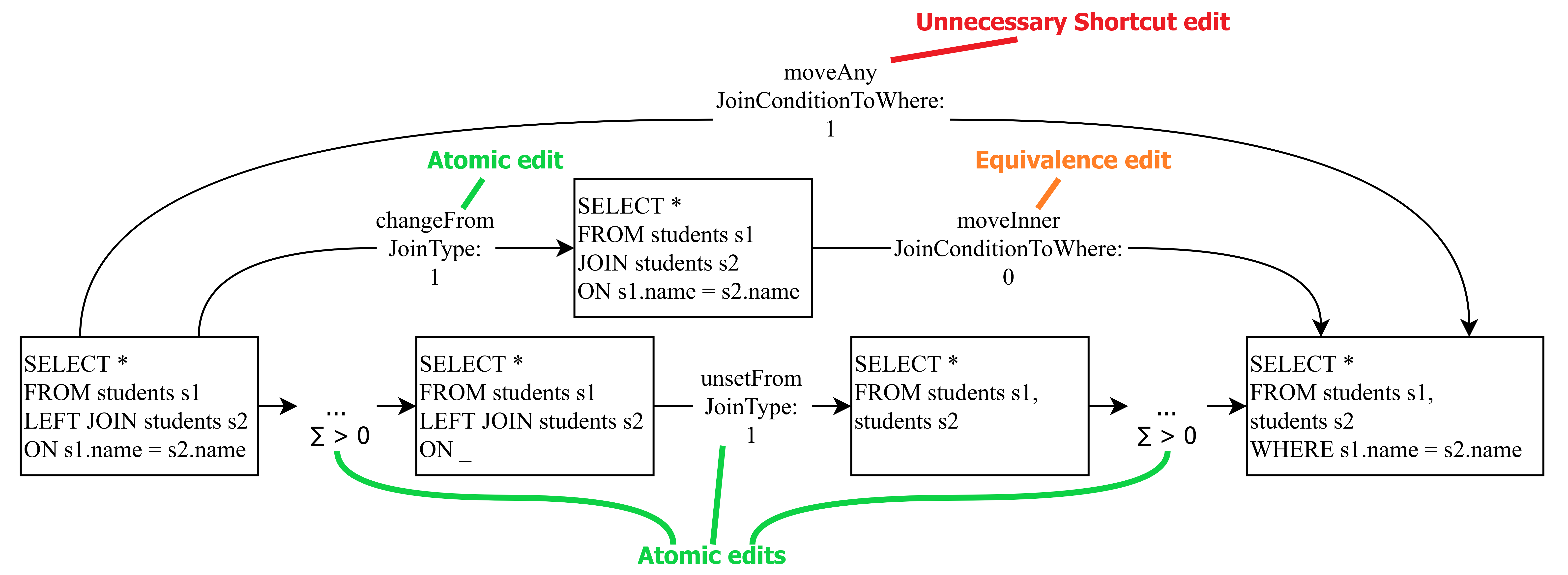}
	\caption{Shortcut edit, that is unnecessary because an equivalence edit combined with an atomic edit already account for it.}
	\Description{Adding shortcut edits only makes sense if they are cheaper than already existing paths.}
	\label{fig:unnecessary-shortcut}
\end{figure}

The question is whether a shortcut edit for the latter case is necessary at all, if there already exists an atomic edit for converting a \texttt{LEFT JOIN} to an \texttt{INNER JOIN}, and a shortcut for relocating the \texttt{INNER JOIN}. \autoref{fig:unnecessary-shortcut} illustrates this.


%% file: src/concept/algorithm.tex
\subsection{Algorithm}\label{sec:algorithm}

We use a shortest path algorithm to find the cheapest edit sequence between a start and target node.
Dijkstra's algorithm, despite being a popular option, cannot be applied to implicit graphs in its commonly taught form. \cite{Felner.2011}
This is why we use uniform cost search as a basis. (See \autoref{sec:uniform-cost-search} for basics.) 

\subsubsection{Search direction}\label{sec:search-direction}

Since our graph is directed, it is necessary to define a fixed direction for the shortest path algorithm. We can start with either the student-devised query or the reference solution.

Advantages for choosing the reference solution as the start and searching towards the student-devised solutions are:
Uniform cost search builds a tree of shortest paths with the start node as the root.
In theory, one run of the shortest path algorithm would be enough to find all the shortest paths to all student-devised solutions.
This can be achieved by comparing every visited node with every (not yet found) solution devised by the students, until there are none left.
If there is a maximum search distance after which a destination is simply considered "too dissimilar", the single execution can even be run in advance to getting student-devised solutions:
For a given start, all visited nodes and edges up to this distance are stored.
Later, a given target can be compared to these nodes.
If none match, it is marked as too dissimilar, otherwise the shortest path can be found by following the tree back to its root.

The disadvantages outweigh these advantages:
%
It seems unrealistic to try to generate every possible student solution, since no assumptions or even requirements can be made.
The presented approach is designed to handle as much input as possible, with the only requirement being that it can be parsed into a (possibly incomplete) \ac{ast}.
The reference solution, on the other hand, can reasonably be required to be executable. This reduces the search space substantially, since no unreasonable values, references or constructs need to be generated, and existing ones can simply be replaced/removed. This includes, for example, references to tables or columns that do not exist in the schema, or illegal expressions such as an aggregation function in the \texttt{WHERE} clause.
For this reason, it is recommended to start with the student-devised input and search towards the reference solution.

\subsubsection{Custom Algorithm}\label{sec:custom-algorithm}

The custom algorithm behind this paper's approach is a variant of uniform cost search that better adapts to and exploits certain properties of the problem.

The most important property is the fixed set of edits and their fixed costs.
Due to this, the distances to all possible neighbors of a node are known in advance.
This information is used to lazily postpone the generation of these neighbors (= execution of the neighbor function) as long as possible: 

When a node \texttt{v} becomes visited, because its distance \texttt{dist[v]} is the smallest of all unvisited nodes, uniform cost search would generate all of its neighbors, filter them for not being visited yet, and then either insert them into the priority queue using \texttt{dist[v]} plus the edge cost as a tentative distance or update their existing tentative distance.
Instead, \texttt{v} itself is inserted into the priority queue at \texttt{dist[v]} plus the cost of the lowest-cost edit and \texttt{next\_edit[v]} is set to this edit.
When a node \texttt{u} is removed from the queue, because \texttt{dist[u]} plus the cost of \texttt{next\_edit[u]} is the smallest queued distance, only \texttt{next\_edit[u]} is applied and each generated neighbor \texttt{n} becomes visited with \texttt{dist[n]} set to the distance that \texttt{u} was queued at.
\textbf{Effectively, not the neighbors are queued, but tuples of a node and edit that will eventually generate them.}
Afterwards, \texttt{next\_edit[u]} is set to the next-higher-cost edit and \texttt{u} is put back into the queue at \texttt{dist[u]} plus the cost of the updated \texttt{next\_edit[u]}.

This way, neighbors are created just before being marked as visited, and \texttt{dist[]} values are always final, but nodes are queued multiple times.
In the worst case, nodes will be visited as often as there are edits, but most will be visited less often.
This is the big advantage of this algorithm, because not doing unnecessary edits also means not having to generate, process, and store unnecessary neighbors.
The resulting algorithm can be described like this:
\begin{enumerate}
    \item Sort all edits by ascending cost into a list called \texttt{edits}.
          The cost of the first edit in this list is called the first edit cost.
    \item If start and destination are equal, return the distance 0.
    \item Mark the start as visited,
          set \texttt{dist[start]} to 0 and \newline\texttt{next\_edit[start]} to the first edit in \texttt{edits},
          and add the start to the min-priority queue, with the first edit cost as priority.
    \item\label{item:loop-start} Extract the lowest-priority entry from the queue as the current node.
          The current distance is \texttt{dist[current\_node]} plus the cost of the edit \texttt{next\_edit[current\_node]}.
          Generate neighbors of the current node by applying the edit \texttt{next\_edit[current\_node]}.
          For each of these neighbors:
          \begin{enumerate}
              \item If the neighbor has already been marked as visited, discard it.
              \item If the neighbor is equal to the destination, return the current distance.
              \item Mark the neighbor as visited,
                    set \texttt{dist[neighbor]} to the current distance and \texttt{next\_edit[neighbor]} to the first edit in \texttt{edits},
                    and add the neighbor to the queue, with \texttt{dist[neighbor]} plus the first edit cost as priority.
          \end{enumerate}
    \item If \texttt{next\_edit[current\_node]} is not yet the last edit in \newline\texttt{edits},
          set it to the subsequent one and add the current node to the queue, with \texttt{dist[current\_node]} plus the cost of \newline\texttt{next\_edit[current\_node]} as priority.
    \item If the queue is empty, the destination cannot be reached from the start, so stop.
          Otherwise, go back to step \ref{item:loop-start}.
\end{enumerate}

To reconstruct the shortest path, every neighbor's predecessor \texttt{prev[neighbor]} and source edit \texttt{edit[neighbor]} can be stored when it becomes visited.
Once the destination has been reached, the path can be reconstructed by reverse-iterating on \texttt{prev[]} from the destination back to the start.

It is also desirable to only search for the destination within a certain distance.
When trying to determine equivalence, it is only of interest whether the destination is within a distance of 0.
For partial grading, it may be unnecessary to keep searching if start and destination are so dissimilar that no points are awarded anyway (see \autoref{sec:result}).
This can be done by aborting as soon as \texttt{current\_distance} is greater than the maximum search distance.

\citeauthor{Chandra.2021} (see \autoref{sec:comparing-for-similarity}) propose using a greedy heuristic algorithm and so-called guided edits. While this reduces the search space and thus improves performance significantly, it can also make the result incorrect, which is why we decided against it.

%% file: src/concept/termination.tex
\subsection{Finite Termination}\label{sec:termination-and-correctness}


It is already proven that if edge weights are greater than or equal to 0, and the target is reached (i.e., visited) by uniform cost search, it is reached by the shortest possible path.~\cite{Felner.2011}
Since the custom algorithm behaves like uniform cost search, except that neighbors are generated as late as possible (see \autoref{sec:custom-algorithm}), it is trivial to apply the same invariants from the proof.

The first condition is that the edge weights are non-negative. This was specified in \autoref{sec:edits-vs-edges}.
The second condition is that the target is reached.
This, in turn, requires that there is a path to find and that the algorithm finds it, which means it must terminate in finite time.

\subsubsection{Existence of a path}\label{sec:existence-of-a-path}


When considering whether a path exists, the atomic edits are essential. Half of the atomic edits are responsible for unsetting attributes, removing (sub)components, etc., and there is such an edit for every possible attribute/(sub)component.
Combined with the support for arbitrary incomplete \acp{ast}, by gradually unsetting and removing everything from the start, the fully incomplete \ac{ast}, called "empty \ac{ast}", can eventually be reached.
And since any given start only has a finite size, this takes a finite number of steps.

The other half of the atomic edits is responsible for setting attributes, adding (sub)components, etc., and there is also an edit for each possible attribute/(sub)component.
Combined with the support for arbitrary incomplete \acp{ast}, by gradually setting and adding everything to the empty \ac{ast}, the target can eventually be reached.
And since any given target has a finite size, this takes a finite number of steps.

Therefore, there exists a path with a finite number of steps from any given start to any given destination.

\subsubsection{Termination of the algorithm}\label{sec:termination-of-algorithm}

If there is at least one path, the algorithm also has to find it in finite time.
The main problem is the graph being implicit and infinitely large.

There are two points with potential for non-termination:
The outermost loop, henceforth called the "main loop", and the generation of neighbors/execution of edits.
Other loops, such as iterating over the set of edits, depend on the size of the input, which must be finite.
Otherwise, passing arguments without even starting the algorithm would take infinitely long.

Termination-related considerations can be reduced to only the main loop:
If the main loop has finite iterations/queue elements to process, then the edits have produced only a finite number of nodes.
Because in order for an edit to produce infinitely many nodes, but the main loop then only having to process a finite subset of them, there are two options:
either the edit would have to generate and discard infinitely many different nodes through the duplicate check, which implies infinitely many nodes have been visited through endless iterations of the main loop, contradicting our assumption; or, the same node would have to be created multiple times by the same edit, which defies the definition of edits returning a set of nodes.
It is therefore sufficient to show that the main loop terminates.

The main loop can only terminate by reaching the goal (excluding the option of terminating early due to exceeding a maximum distance). 
This is because there always is at least one path, so it is impossible to run out of neighbors before reaching the goal.
The danger of non-finite termination comes from having to process infinitely many nodes before this happens.
For example, if there are infinitely many nodes with distance 0, but the destination has a distance of 1, it will never be reached.
Since nodes are processed in the order of increasing distance from the start, this condition can be weakened to there being a finite number of nodes at all distances smaller than or equal to the distance to the destination.

\begin{figure}[htb]
    \centering
    \includegraphics[width=0.9\linewidth]{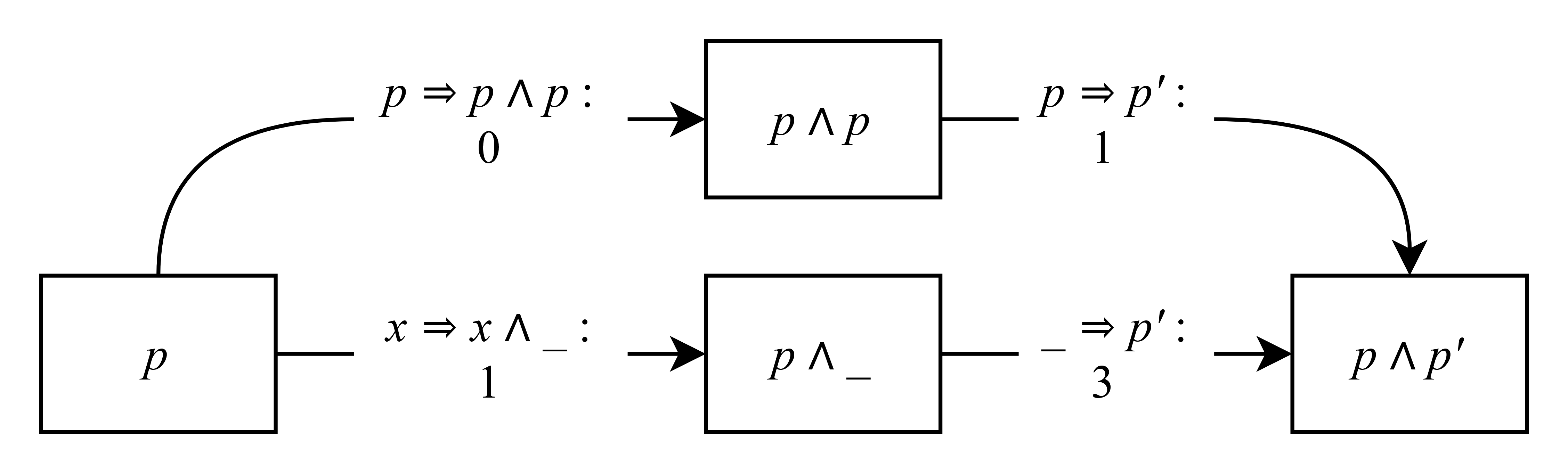}
    \caption{The equivalence edit "$p \Rightarrow p \wedge p$" and transformation "$p \Rightarrow p'$" being cheaper than introducing the binary expression "$\wedge$" and creating "$p'$" from scratch.}
    \Description{This cheaper path contains an equivalence edit with cost 0 that increases the \ac{ast} size.}
    \label{fig:equivalence-edit-problem}
\end{figure}

This is ensured individually for certain portions of edits:
\begin{description}
    \item[Atomic unset- and remove-edits:]
          The amount of nodes possibly generated by atomic unset- and remove-edits is finite, because it is limited by the components present in the \ac{ast}.

    \item[Atomic set-edits:]
          For atomic set-edits, the amount is upper-bound by the number of possible values to the power of the number of attribute-places: $amount \leq {values}^{places}$.
          The places are limited because of the finite \ac{ast} size.
          But the number of values, especially for literals/constants, is not.
          A finite set of values must be determined based on those in the destination and the current node.
          Such information is passed to the edit as part of the meta-info argument.

    \item[Atomic add-edits:]
          For atomic add-edits, the amount is upper-bound by the number of different components to the power of the number of places to add them times the number of executions: $amount \leq {components}^{({places} \times 2^{executions})}$.
          The power of 2 is due to components being nestable, i.e. adding a binary-expression creates 2 new places to add expressions.
          The number of different components is limited by the syntax.
          The places to add them are limited because of the finite input size.
          And because atomic add-edits generally have a cost greater than 0, the number of executions up to a limited distance is also limited, making the amount finite.
          \par
          However, the cost of any edit is configurable to 0.
          This allows for unlimited executions by growing the \ac{ast} infinitely large.
          To solve this, the size of the \ac{ast} is limited as described further below.

    \item[Horizontal equivalence edits:]
          Equivalence edits, that don't change the components but only their positions, e.g., $p~\wedge~q \Leftrightarrow q \wedge p$, are innately limited by the number of possible combinations.
          Those, that change attributes, are limited by the same finite set of values as for atomic set-edits.

    \item[Equivalence edits, that don't increase the \ac{ast} size:]
          Equi-valence edits, that eliminate or change components, but don't increase the \ac{ast} size, are limited in the same way the atomic remove-, unset-, and set-edits are.

    \item[Equivalence edits, that do increase the \ac{ast} size:]
          Critical \\are equivalence edits, that do increase the \ac{ast} size.
          In particular if their output contains the input again, e.g., the right direction of $p \Leftrightarrow p \wedge p$, then they can be chained infinitely, like $p \Leftrightarrow p \wedge p \Leftrightarrow p \wedge p \wedge p \Leftrightarrow p \wedge p \wedge p \wedge p \Leftrightarrow ...$ .
          \par
          Applying these equivalences only in the simplifying direction, e.g., only the left direction of $p \Leftrightarrow p \wedge p$, would break correctness.
          For example, assume the destination contains $p \wedge p'$, with $p$ being similar to $p'$, such that transforming the former into the latter is cheaper than creating $p'$ from scratch. 
          If the start now only contains $p$, then using the right direction of $p \Leftrightarrow p \wedge p$ to turn it into $p \wedge p$, followed by said transformation to get $p \wedge p'$, is cheaper than the alternative (see \autoref{fig:equivalence-edit-problem}).
          Applying 
          transformations on the destination instead, e.g., $p'$ to $p$ and then $p \wedge p$ to $p$, to cover such cases, has been ruled out in \autoref{sec:search-direction}.
          \par
          Equivalence edits, that increase the \ac{ast} size, are instead limited by limiting the \ac{ast} size itself.
          This can be done via the number of components, the height and degree of the tree, or both.
          The information is passed to edits as part of the meta-info argument.
          The exact values depend on the destination and all edits, because it has to be assured that no shortest path is obstructed. 
          Since custom edits might be involved, this cannot be further determined here.
\end{description}

This way, the number of nodes with a distance smaller than or equal to that of the destination is finite, so the algorithm reaches the destination and terminates in finite time.

%% file: src/evaluation.tex
\section{Evaluation}\label{sec:evaluation}

Our approach has been evaluated to see if it theoretically fulfills the requirements of \autoref{sec:goals}.
In addition, a prototype implementation was created and a survey was conducted to see if it could be successfully implemented in practice with respect to fairness and comprehensibility.

\subsection{Conceptual evaluation}\label{sec:conceptual-evaluation}

The concept from \autoref{sec:concept} delivers on all goals set in \autoref{sec:goals}:


\subsubsection{\hyperref[item:goal-quantify]{Quantified semantic similarity}}\label{sec:evaluation-quantify}

The primary goal is satisfied by the underlying idea specified in \autoref{sec:idea}.
Queries are treated like nodes in a graph.
The edges between them are implicitly given by edits to those queries.
Edits, and therefore edges, have costs associated with them that represent the semantic error or dissimilarity caused by them.
On this graph, a shortest path algorithm can be used to find the lowest cost edit sequence between any two nodes (see \autoref{sec:termination-and-correctness}).
The sum of the costs of these edits represents how semantically dissimilar the queries are.
A cost of 0 means that the queries are considered to be semantically equivalent.
Otherwise, the higher the value, the less similar they are.

\subsubsection{\hyperref[item:goal-feedback]{Meaningful feedback}}\label{sec:evaluation-feedback}

Instead of combining all edits by their cost, they are considered individually.
Each edit can be described in natural language.
Listing the descriptions of the edges along the path serves as the required feedback, because they either prove why the queries are semantically equivalent or they explain what would be needed in order to make them equivalent.





\subsubsection{\hyperref[item:goal-result]{Guaranteed result}}\label{sec:evaluation-result}

We prove in \autoref{sec:termination-and-correctness} that the approach meets the goal of always yielding a result.
The atomic edits ensure that there is a path, while the number of nodes to be visited before reaching the destination is finite.



\subsubsection{\hyperref[item:goal-input]{Unrestricted input}}\label{sec:evaluation-input}

The \acp{ast}, which form the nodes of the graph, can represent any \ac{sql} construct.
In addition, they can represent construct or value combinations that are non-executable or even incomplete.
While it makes sense to restrict the target to executable queries to limit the search space (see \autoref{sec:search-direction}), the start is not restricted in this way.

\subsubsection{\hyperref[item:goal-configure]{Configurability and extensibility}}\label{sec:evaluation-configure}

The cost of any edit can be configured individually.
This enables the approach to adapt to any teaching style or priorities.
Further, even custom edits can be added to account for task-specific cases.

\subsection{Prototype implementation}\label{sec:implementation}

In order to prove that the presented approach works in practice, a prototype implementation has been created (\url{https://github.com/FAU-CS6/sql-query-distance}).

It was written in TypeScript to eventually be used in a browser-based plugin.
At the time of writing, it features 181 edits, 94 of which are the atomic edits.
The rest of them are the shortcut edits that enable detection of semantical equivalence and improve the quality of the calculated distance in general.
All of them have descriptions in natural language used to provide meaningful feedback.
Also, they have adjustable costs and are extensible.


\subsection{Survey}\label{sec:survey}

We conducted a survey to evaluate our approach by comparing it against two other techniques, dynamic analysis and manual grading, in terms of two metrics, fairness and comprehensibility.

The survey consisted of 11 scenarios.
Each one began with a task description like typically found in \ac{sql} exercises or exams:
first a database schema is presented and then the student is asked to extract specific data from said database by writing a corresponding \ac{sql} query.

Next, the scenario presented the reference solution and a student-devised answer.
To make the survey as realistic as possible, the latter consisted only of real answers from real students.
The data was collected by an automated grading system based on dynamic analysis offered to the students for autonomous practice and was anonymized.

\begin{figure}[htb]
    \centering
    \begin{subfigure}[]{\linewidth}
        \centering
        \includegraphics[width=\linewidth]{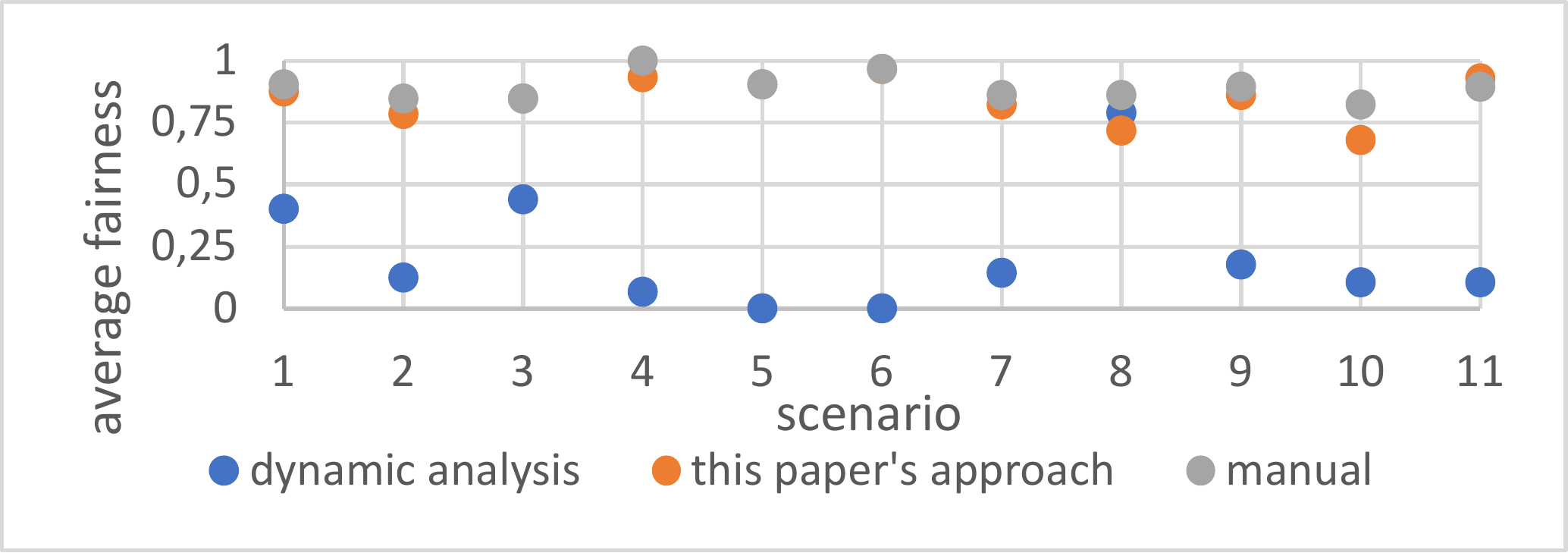}
        \caption{The average fairness of each technique per scenario.}
        \Description{Dynamic analysis mostly performs quite poorly while this paper's approach consistently scores almost as well as manual grading.}
        \label{fig:survey-fairness}
    \end{subfigure}
    \begin{subfigure}[]{\linewidth}
        \centering
        \includegraphics[width=\linewidth]{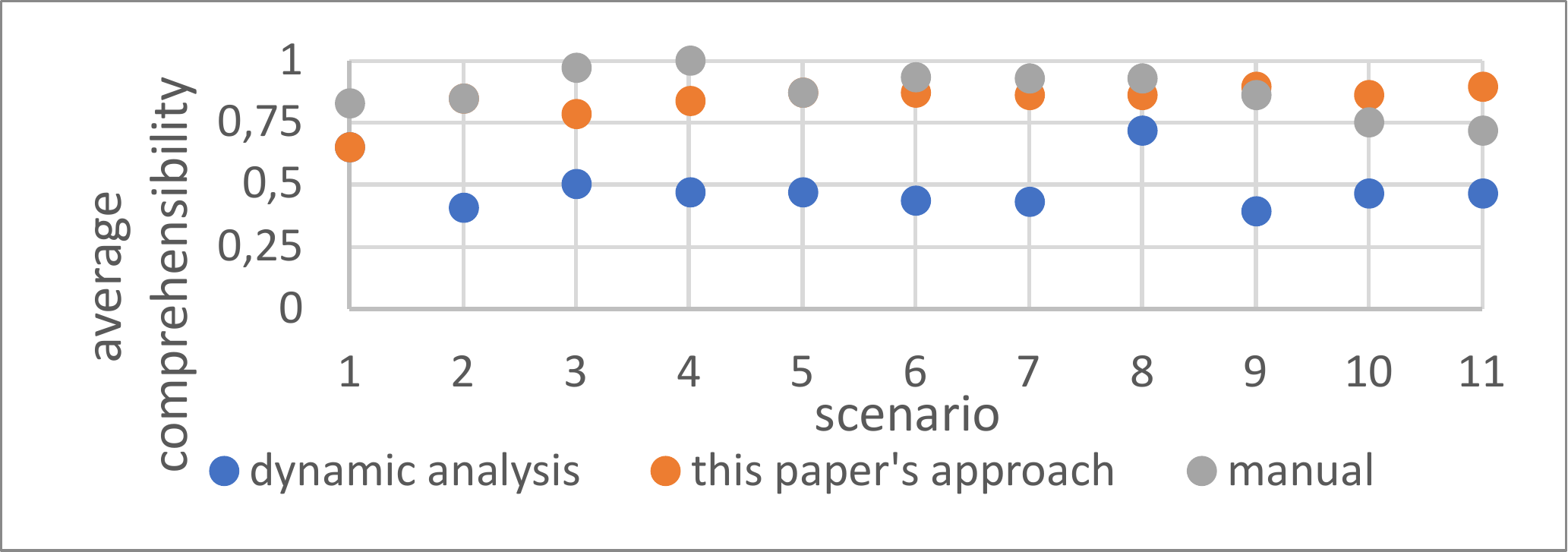}
        \caption{The average comprehensibility of each technique per scenario.}
        \Description{Dynamic analysis again mostly performs poorly while this paper's approach consistently scores almost as well, sometimes even better than manual grading.}
        \label{fig:survey-comprehensibility}
    \end{subfigure}
    \caption{The summarized survey results per scenario.}
    \Description{Two subfigures containing a scatter plot, each, of the three techniques' average fairness and comprehensibility, respectively.}
    \label{fig:survey-results}
\end{figure}

The scenario finally listed the grading given by the system based on dynamic analysis at that time, as well as a grading computed from query distance according to our prototypical implementation and a manually assigned grading.
The survey then asked participants to rate the 3 gradings in this scenario by fairness and understandability.
The possible answers "low", "medium", and "high" were linearly mapped to a scale from 0 to 1 to facilitate quantitative analysis.
The results can be found in \autoref{tab:survey-results} and \autoref{fig:survey-results}.

\begin{table}[htb]
    \begin{tabular}{@{}lllll@{}}
        \toprule
        technique             & \multicolumn{1}{c}{$\bar{f}$} & \multicolumn{1}{c}{$s_f$} & \multicolumn{1}{c}{$\bar{c}$} & \multicolumn{1}{c}{$s_c$} \\ \midrule
        dynamic analysis      & $0,2186$                      & $0,3611$                  & $0,4940$                      & $0,4238$                  \\
        this paper's approach & $0,8473$                      & $0,2608$                  & $0,8293$                      & $0,2935$                  \\
        manual                & $0,8892$                      & $0,2283$                  & $0,8743$                      & $0,2608$                  \\ \bottomrule
    \end{tabular}
    \caption{The overall sample means of fairness $\bar{f}$ and comprehensibility $\bar{c}$, as well as their (uncorrected) sample standard deviations $s_f$ and $s_c$, respectively, for each technique.}
    \label{tab:survey-results}
\end{table}

\autoref{tab:survey-results} summarizes the answers from all 20 participants:
Overall, the fairness of this paper's approach was rated $287.7\%$ higher than dynamic analysis and just $4.2\%$ lower than manual grading.
Also, it is shown to be $1.68$ times as comprehensible as dynamic analysis, or $94.9\%$ as much as manual grading.

\autoref{fig:survey-fairness} shows that, across all scenarios, dynamic analysis mostly performs quite poorly when it comes to fairness, while this paper's approach consistently scores almost as well as manual grading, if not slightly better.
This means that even in a prototypical stage, the implementation from \autoref{sec:implementation} can already compete with manual grading in terms of perceived fairness, indicating a high quality of query comparison.

\autoref{fig:survey-comprehensibility} paints a similar picture regarding comprehensibility, but with this paper's distance-based approach even significantly outperforming manual grading in the last two scenarios.
This stands testament to the powerful semantic feedback from \autoref{sec:evaluation-feedback}.

Scenario 8 being an outlier in both metrics stems from an ambiguous task description, which later turned out to even divide our team as to what should be considered a "correct" answer.
Since this is an issue separate from query comparison techniques, it will not be investigated further.

%% file: src/future-work.tex
\section{Discussion}\label{sec:discussion}

Our prototypical implementation provides only a subset of edits and does not yet support all \ac{sql} features. While the set of equivalence edits can never be fully exhaustive, it makes sense to extend them as much as possible, as this will automatically improve the accuracy for non-equivalent queries as well. Adding support for the missing \ac{sql} features is just a matter of time and effort. 

Another problem with the prototypical implementation is the performance for complex queries, especially if they have a large semantic distance, since the search space grows exponentially. We avoided canonicalization as part of the node comparison because of the risk of removing important syntactic distinctions, but didn't introduce an equally good way to save performance. There is great potential for parallelization, especially when applying edits, which is untapped in our TypeScript-based implementation.

A possible performance improvement might be to use a different shortest path algorithm. 
For example, one could use only the atomic edits to quickly build an in initial path, which gives an upper bound for the distance. Then, shortcut edits could be used to continuously search for cheaper variations. This converges towards the correct result, but can be aborted early to get an approximated, but still valid, path.

While our approach can handle non-executable queries, it cannot handle unparsable ones.
However, this could be achieved by combining it with other comparison techniques.
For example, string similarity, as suggested by \citet{Wang.2020}, could be used as a backup solution for query descriptions that cannot even be parsed into an incomplete \ac{ast}.
In this case, however, care must be taken to ensure that such backup solutions never produce better grades than the base comparison technique, as this might incentivize students to deliberately produce "broken" answers in order to get a better grade.

A comparison with \citeauthor{Chandra.2021} as part of our survey was planned, but couldn't be implemented due to technical problems while trying to deploy their implementation. As they are the closest to our approach, this would have been a very interesting comparison.
Also, we have to admit that the number of 20 participants in our survey is comparatively small and there is no way to prove the representativeness of the study group. However, due to the visible trends across the different scenarios, we believe that this survey is at least a strong indicator in favor of the presented approach.

%% file: src/conclusion.tex
\section{Conclusion}\label{sec:conclusion}

Our goal in this paper was to develop a method for quantifying the semantic similarity of two queries in a way that is configurable and extensible, while always yielding a result and not limiting input to a small subset of the query language. If needed, the method also provides meaningful feedback to the user.

We proposed a graph-based approach that treats queries as nodes and edits as edges. The lowest-cost sequence of edits required to transform one query into another is then determined by applying a shortest path algorithm. The sum of the costs of these edits is the dissimilarity of the queries. The edits are configurable and extensible, making the approach adjustable to different contexts and points of focus. The approach is not limited to a subset of the query language, and can even be extended by incomplete ASTs.

We proved finite termination of the approach and showed its feasibility and practicality by implementing a prototype for the comparison of student SQL queries against a reference solution and conducting a survey on its perceived fairness and comprehensibility. We are confident that our approach can be applied to other domains as well, and that it can be extended to support more complex queries and other query languages.

Although the results of the survey have already been very promising, we plan to refine the prototype for enhanced \ac{sql} support and robustness against incomplete queries. Additionally, the prototype will be integrated into an automated grading system, providing a long-term test beyond our initial survey.